\documentstyle[twocolumn,aps]{revtex}
\begin{document}
\draft
\title{Oscillating magnetoresistance in diluted magnetic semiconductor barrier
structures }
\author{Kai Chang \cite{addnow} and J. B. Xia}
\address{NLSM, Institute of Semiconductors, Chinese Academy of Sciences,\\
Beijing 100083, China}
\author{F. M. Peeters \cite{addnew}}
\address{Department of Physics, University of Antwerp (UIA), Universiteitsplein 1,\\
B-2610 Antwerpen, Belgium}
\date{\today}
\maketitle

\begin{abstract}
Ballistic spin polarized transport through diluted magnetic semiconductor
(DMS) single and double barrier structures\ is investigated theoretically
using a two-component model. The tunneling magnetoresistance (TMR) of the
system exhibits oscillating behavior when the magnetic field are varied. An
interesting beat pattern in the TMR and spin polarization is found for
different NMS/DMS double barrier structures which arises from an interplay
between the spin-up and spin-down electron channels which are splitted by
the s-d exchange interaction.
\end{abstract}


Spin polarized transport in solid-state systems has generated intense
interest as it is the crucial ingredient for spintronics\cite{Datta,SPT} and
several quantum computation scheme\cite{Kane,Loss}. For these applications,
a basic requirement is to produce, and sustain, high spin-polarized currents
in semiconductors for sufficient long times. Several schemes have been
proposed to produce spin injection in semiconductors, such as electrons
injected from a ferromagnetic metal into a semiconductor \cite
{Lunn,Lee,Hammar}, but the change in device resistance for parallel and
antiparallel magnetization is very small. Diluted magnetic semiconductors
(DMS) provides us with a new system in which spin dependent optical and
transport properties are expected. The spin-dependent transport and optical
properties in DMS\ systems arise from the {\it s-d} exchange interaction
between the conduction electron and the localized $3d^5$ electrons of the Mn
ions which lifts the degeneracy of the spin-up and spin-down electron and
hole states.\cite{Kossut}

Spin dependent optical properties in a DMS spin superlattice was proposed 
\cite{Ortenberg} and realized\cite{Dai}. Spin-dependent tunneling through a
DMS\ junction was also studied theoretically within a mean-field
approximation \cite{Egues,Egues1,Chang}. Very recently, experiments have
demonstrated that a robust spin injection through a diluted magnetic
semiconductor (DMS) junction Be$_x$Mn$_y$Zn$_{1-x-y}$Se \cite{Fiederling},
and a Mn doped p-type\ GaAs spin aligner was possible in which holes were
injected into GaAs in the presence of an in-plane magnetic field \cite{Ohno}%
. Spin polarization of the injected carriers was detected by the emitted
circular polarized light from the holes which recombined with the electrons
in the nonmagnetic semiconductor (NMS) quantum wells. Spin coherence can be
maintained\ in semiconductors over large distances ($\geq 100\mu m$) and for
long time (up to nanoseconds) \cite{Kikkawa}.

In this work, we report an interesting oscillating tunneling
magnetoresistance (TMR) and spin polarization (SP) through NMS/DMS
structures. The DMS structure is similar to the sample used in the
magneto-optical study of Ref. \cite{Dai}. We find theoretically that the TMR
for the double barrier structure oscillates with increasing magnetic field
and exhibits a peculiar beat pattern. The underlying physics of the
phenomena arises from the interplay of the spin-up and the spin-down
channels which are split by the {\it s-d} exchange interaction. Note that
band-structure effects and spin-orbit interaction are not very efficient
spin-flip processes for electrons.\cite{Bastard} Recently, Egues {\it et al} 
\cite{Egues1} presented results on spin filtering and magnetoresistance
through ballistic tunneling junctions. These results are complementary to
ours, i.e. the beat pattern in the double NMS\ barriers with DMS\ contacts
is similar to what we found, but we additionally studied double DMS barriers
with NMS\ contacts which we found to exchibit also beat patterns.
Furthermore, in our calculation we used a different approach which is more
easily applicable to more complicated cases, such as e.g. the case with an
electric bias. In this work we prove that the beat pattern which arises from
the superposition of the spin-up and the spin-down channel is a rather
general phenomena and can be observed in either NMS\ double barrier
structures with DMS\ contacts or DMS barriers with NMS\ contacts. Instead of
ZnSe/Zn$_{1-x}$Mn$_x$Se structures used in the calculation of Ref. \cite
{Egues1}, we propose Cd$_{{1-y}}$Mg$_y$Te / Cd$_{{1-x}}$Mn$_x$Te structures
to realize single and double NMS(Cd$_{{1-y}}$Mg$_y$Te) barrier structures
with DMS\ contacts(Cd$_{{1-x}}$Mn$_x$Te).

Consider a spin unpolarized electron injected into a NSM/DMS/NSM single or
double barrier structure in the presence of a perpendicular magnetic field.
Due to the s-d exchange interaction, an external magnetic field gives rise
to a giant Zeeman splitting of the conduction band states which results in a
striking difference of the potential profiles seen by the spin-up and the
spin-down\ electron (see the insets of the figures)\cite{Egues}. In this
simple system, the electric current has contributions from spin-up and
spin-down channels. Since the sample dimensions are much smaller than the
spin coherence length, which may reach up to 100$\mu $m in semiconductors 
\cite{Fiederling,Kikkawa}, spin-flip processes can be neglected in our
system.

The model Hamiltonian of such systems is of the following form 
\begin{eqnarray}
H &=&({\bf p+}e{\bf A)}^2/2m+V(z)+1/2g_s\mu _B{\bf s\cdot B}  \label{Hamil}
\\
&&+J_{s-d}\sum_i{\bf s}({\bf r}){\bf \cdot S}({\bf R}_i)\delta ({\bf r}-{\bf %
R}_i)\ ,  \nonumber
\end{eqnarray}
where ${\bf S}$ is the spin of the localized $3d^5$ electrons of the Mn ions
with $\ S=5/2$ and ${\bf s}$ is the electron spin. Here we assume that the
magnetic ions are distributed homogeneously in the DMS layers. $m_e^{*}$ is
the electron effective mass, $V(z)$ is the zero magnetic field potential
profile of, e.g. ZnSe/Zn$_{1-x}$Mn$_x$Se DMS double barrier structures\cite
{Dai}.$\ $The vector potential is taken as ${\bf A=(-}By/2,Bx/2,0{\bf )}$
and therefore the magnetic field points along the $z$-axis. The third term
describes the Zeeman splitting of the electron. The last term in Eq. (1)
denotes the {\it s-d}\ exchange interaction between the electron and Mn
ions. $<S_z>=SB_J(Sg\mu _BB/k_B(T+T_0))\ $, where $B_J(x)$ is the Brillouin
function and $S=5/2$ spins of the localized $3d^5$ electrons of the Mn
ions,\ $N_0$ is the number of cations per unit volume, $J_{s-d}=-N_0\alpha
x_{eff}$ denotes the exchange integral for the conduction band, $\Delta
_Z=J_{s-d}<S_z>$ is the giant Zeeman splitting. The phenomenological
parameters $x_{eff}$ (reduced effective concentration of Mn) and $T_0$
account for the reduced single-ion contribution due to the antiferromagnetic
Mn{\it -}Mn coupling, and ${\bf s}$ is the electron spin operator. The third
and fourth terms in Eq. (\ref{Hamil}) can be viewed as an effective
potential which is different for the spin-up and spin-down electrons (see
the inset of Fig. 1). The parameters used in our calculation are taken from
Ref.\cite{Dai} for ZnSe/Zn$_{1-x}$Mn$_x$Se (see Fig. 1 (x=0.07) and Fig. 4
(x=0.2)) and from Ref. \cite{Ossau} for Cd$_{{1-y}}$Mg$_y$Te/Cd$_{{1-x}}$Mn$%
_x$Te (see Fig. 2 (x=0.04, y=0.08) and Fig. 3 (x=0.04, y=0.05)). The
parameters used in our calculation are as follows: m$_e^{*}$=0.16, T$_0$%
=1.4K, $N_0\alpha $=0.27eV, V$_b\approx $10meV for Zn$_{0.93}$Mn$_{0.07}$Se
and V$_b\approx $75meV for Zn$_{0.8}$Mn$_{0.2}$Se; m$_e^{*}$=0.096, T$_0$%
=3.1K, $N_0\alpha $=0.22eV, V$_b\approx $10meV for Cd$_{0.92}$Mg$_{0.08}$%
Te/Cd$_{0.96}$Mn$_{0.04}$Te and V$_b\approx $75meV for Cd$_{{0.95}}$Mg$%
_{0.05}$Te/Cd$_{0.96}$Mn$_{0.04}$Te.

Electrons on the left and right hand side of the tunneling barrier can be
expressed as $\psi _L^\sigma =e^{ikz}\chi _\sigma +r_\sigma e^{-ikz}\chi
_\sigma $ and $\psi _R^\sigma =t_\sigma e^{ikz}\chi _\sigma $, respectively.
where the spinor $\chi _\sigma (\sigma =\uparrow \downarrow )$ is the
spin-up $\left| \uparrow \right\rangle $ or the spin-down $\left| \downarrow
\right\rangle $ state, $r_\sigma (t_\sigma )$ is the reflection
(transmission) coefficient respectively. Taking into account the boundary
conditions i.e. the continuity of the envelope function $\psi _i^\sigma $
and its derivative $(\psi _i^\sigma )^{^{\prime }}/m_i$ at the interface, we
can connect the electron wave functions $\psi _L^\sigma =T$ $\psi _R^\sigma $
at the boundary, where $T$ is the transfer matrix which has the form $%
T=\prod_j T_M^j=T_M^nT_M^{n-1}T_M^{n-2}\cdot \cdot \cdot T_M^2T_M^1$, with

\begin{equation}
T_M^j=\left( 
\begin{array}{cc}
\cos k_ja & \frac{m_j}{k_j}\sin k_ja \\ 
-\frac{k_j}{m_j}\sin k_ja & \cos k_ja
\end{array}
\right) ,  \label{TM}
\end{equation}
where $k_j=\sqrt{2m_e(E-V_j)/\hbar ^2}$.

When a small bias is applied across the junction, a nonequilibrium electron
population will be generated. The current density $J=\sum_\sigma J^\sigma $
can be calculated \cite{Blonder} 
\begin{equation}
J^\sigma =\frac{e^2V}{2\pi ^2l_B^2}\sum_n(\frac 1{2\pi })\int_0^{k_n^F}dk_z(%
{\bf -}\frac{\partial f_0}{\partial E})T^\sigma v_z^\sigma ,\ \ \ \ \ \ 
\end{equation}
where $T^\sigma (n,E_F)$ is the transmission coefficient of our tunnel
structure at the Fermi surface for the different spin orientations, $%
v_z^\sigma =\hbar k_z^\sigma /m$ is \ the group velocity, $1/2\pi l_B^2$ is
the density of state of each Landau level, $l_B=\sqrt{\hbar /eB}$ is the
magnetic length,$\ \ f_0$ is the equilibrium distribution function of the
conduction band electron, and we use the approximation ${\bf -}\frac{%
\partial f_0}{\partial E}\approx \delta (E-E_F)$ which is valid for $k_BT\ll
E_F$\ . The low-temperature conductance is given by\cite{Chang} 
\begin{equation}
\sigma ^{\uparrow \downarrow }/\sigma _0=\sum_nT^{\uparrow \downarrow
}(k_{nF}^{\uparrow \downarrow }),  \label{conduc}
\end{equation}
where $k_{nF}^{\uparrow \downarrow }=\sqrt{2m/\hbar ^2(E_F-E_n)}$, $\sigma
_0=e^2/2\pi l_B^2h$, $E_F^{\uparrow \downarrow }=E_F\pm \Delta _Z$, $%
E_n=(n+1/2)\hbar \omega _c$ is the energy of the Landau level. In our
formalism, the total conductivity is the sum of the conductivity of each
Landau level at the Fermi surface and this for each spin state.

The degree of spin polarization (SP) of the current density is defined by 
\begin{equation}
P=\frac{J^{\downarrow }-J^{\uparrow }}{J^{\downarrow }+J^{\uparrow }}.
\end{equation}
Here $J^{\uparrow }(J^{\downarrow })$ is the spin-up (spin-down) current
density of the spin-polarized current.

The magnetoresistance (TMR) as a result of tunneling through the NMR/DMS
structures is defined by 
\begin{equation}
\Delta R/R=\frac{R(B)-R(0)}{R(0)}=\frac{R(B)}{R(0)}-1=\frac{\sigma (0)}{%
\sigma (B)}-1.
\end{equation}
Here $J^{\uparrow }(J^{\downarrow })$ is the spin-up (spin-down) current
density of the spin-polarized current.

Fig. 1 depicts how the TMR $\Delta $R/R varies with magnetic field in a DMS
single barrier structures with NMS contacts for different thicknesses of the
DMS layer. The inset shows the spin polarization $P$(SP) versus magnetic
fields. From this figure, we find that the TMR, on the average, decreases
and oscillates with increasing magnetic field. The oscillations of the TMR
and the SP are weakened by increasing the DMS\ barrier thickness. These
oscillations are mainly attributed to the oscillation of the spin-down
conductivity component $\sigma ^{\downarrow }/\sigma _0$ which is enhanced
(weakened) by increasing magnetic field (the thickness of the DMS layer).
The conductivity $\sigma ^{\downarrow }/\sigma _0$ of the spin-down
component is larger than that of the spin-up component, since the barrier
height seen by the spin-up electron is higher than that seen by the
spin-down electron due to the magnetic field-induced {\it s-d} exchange
interaction. Therefore the spin-polarization increases and saturates with
increasing magnetic field and barrier thickness.

In Fig. 2 we plot the TMR $\Delta $R/R as a function of magnetic field in a
NMS\ single barrier structure with DMS contacts. There is a large difference
between Figs. 1 and 2. Notice the almost step-like character in the TMR and
the SP when the magnetic field approach the critical field $B_c\sim 4$T
which is determined by the energy separation between the spin-up electron
energy and the Fermi energy. This separation is determined by the magnetic
field and the temperature (see Eq. (\ref{Hamil})) and is independent of the
thickness of the barrier. This step-like behavior arises from the
competition of the spin-up and the spin-down electron conductivity. 
%The magnetic field raises (lowers) the energy of the injected spin-up
%(spin-down) electron states relative to the Fermi energy.
The oscillating conductivity of the spin-up (spin-down) electron decreases
(increases) with increasing magnetic field, the spin-down channel make
dominant contribution to the conductivity and the spin-up channel is blocked
when the magnetic field is larger than the critical field {\it B}$_c$ and
consequently leads to a step. In the inset of Fig. 2,\ the spin polarization
also exhibits a step-like behavior versus the magnetic field.

In Fig. 3 (a), we find that the tunneling magnetoresistance (TMR) $\Delta
R/R $ of a NMS double barrier (DB) structure with DMS contacts oscillates as
a function of the magnetic field. A peculiar beat pattern is clearly seen in
the TMR. A similar behavior is also observed in the spin polarization (see
the inset of Fig. 3(a)). The oscillating behavior arises from the interplay
between the Fermi surface and the position of the Landau levels of the two
spin states. When the magnetic field increases, the Landau levels are swept
across the Fermi surface one by one resulting in oscillations in the
magnetoresistance whose origin is similar to those of Shubnikov de Haas
(SdH) oscillations. In contrast to the usual SdH oscillations the {\it s-d}
exchange interaction leads to a giant Zeeman splitting at low temperature,
and this splitting increases with increasing magnetic field which saturates
for strong magnetic field which leads to an usual pattern of oscillations.
The beating is a result of the fact that the total current is composed of
spin-up and spin-down components which are split by the {\it s-d} exchange
interaction. The interplay between the spin-up and spin-down channels
results in the beat pattern in the magnetoresistance. This is clearly
demonstrated in Fig. 3(b) where we show the total conductivity (the thick
curve) and the spin-up (dotted) and spin-down (dashed) components. The phase
difference of the oscillating TMR of the spin-up and spin-down channels
varies with increasing magnetic field, therefore the reduction and
enhancement of the oscillating conductivity of the spin-up and spin-down
channels lead to the beat pattern of the total TMR through the NMS double
barriers with DMS contacts.

Figures 4(a) and 4(b) show the TMR $\Delta R/R$ as a function of magnetic
field for DMS\ double barriers with NMS contacts. The inset of Fig. 4(a)
shows the spin polarization versus the magnetic field. Similar oscillating
and beat behaviors as in Fig. 3(a) can be found in this figure. Comparing
these results with Figs. 3(a) and 3(b), the beat patterns in the TMR and the
SP are weakened since the phase difference between the spin-up and the
spin-down channel varies slowly with increasing magnetic field for the DMS
double barriers with NMS\ contacts.

In conclusion, we studied theoretically the {\it s-d} exchange interaction
in DMS\ single and double barrier structures. Our theoretical results
demonstrate that the oscillating TMR in DMS\ double barrier structures can
be controlled by an external magnetic field. The beat pattern in TMR arises
from the interference between the spin-up and spin-down channels. The spin
polarization also exhibits an oscillating behavior when the\ thickness of
the DMS layer changes. Notice that the NMS/DMS structures used in our
calculation are already realized experimentally in recent magneto-optical
studies, but at present no transport measurements on such structures are
available. Our results clearly illustrate that the spin polarization of the
tunneling current can be tuned in magnitude and sign by changing the
external magnetic field and/or the width of the tunneling barrier. Such
systems are extremely attractive from the point of view of both basic
research and technological applications, such as, e.g. in spin switches and
spin transistors.

\begin{acknowledgments}
This work was financially supported by CAS, the Chinese Science Foundation, 
the Flemish Science Foundation (FWO-Vl), the Belgian Inter-University Attraction Poles
program (IUAP), the Concerted Action program (GOA), the Inter-University Microelectronics
Center (IMEC, vzw), and the Flemish-Chinese 
bilateral science and technological cooperation. 
\end{acknowledgments}

\begin{figure}[tbp]
\caption{The TMR $\Delta R/R$ as a function of the magnetic field for
different DMS barrier thicknesses. The inset gives the spin polarization
versus the magnetic field. The potential profiles for $B=0$ and $B\ne 0$ are
also plotted in the figure. The solid, dashed, dotted and dash-dotted curves
correspond to the different thicknesses of the DMS layer: 5nm, 10nm, 15nm,
20nm, respectively. The parameters are taken from Ref. \protect\cite{Dai}
for ZnSe/Zn$_{1-x}$Mn$_x$Se (x=0.07)). }
\end{figure}

\begin{figure}[tbp]
\caption{The same as Fig. 1 but now for the DMS barrier with NMS contacts.
The parameters are taken from Ref. \protect\cite{Ossau} for Cd$_{1-y}$Mg$%
_{y} $Te/Cd$_{1-x}$Mn$_{x}$Te (x=0.04, y=0.08)). }
\end{figure}

\begin{figure}[tbp]
\caption{(a) The TMR $\Delta R/R$ as a function of magnetic field for a NMS
double barrier structure with DMS contacts. The inset shows the spin
polarization for $E_F=60meV$, $V_b=75meV$. (b) The total conductivity (thick
solid curve) and the spin-up (dashed curve) and spin-down (dotted curve)
conductivities as a function of magnetic field. The parameters are taken
from Ref. \protect\cite{Ossau} for Cd$_{1-y}$Mg$_{y}$Te/Cd$_{1-x}$Mn$_{x}$Te
(x=0.04, y=0.08)). }
\end{figure}

\begin{figure}[tbp]
\caption{ The same as Fig. 3 (a) and (b) but now for the DMS double barrier
structure with NMS contacts. The parameters are taken from Ref. \protect\cite
{Dai} for ZnSe/Zn$_{1-x}$Mn$_{x}$Se (x=0.2)). }
\end{figure}

\end{document}